\documentclass[11pt]{article}
\usepackage{amsmath,amsthm}
\usepackage{amssymb, amsfonts}
\usepackage{graphicx, natbib,bbm,bm}
\usepackage{enumerate}

\usepackage{lipsum}
\usepackage{hyperref}

\newtheorem{theorem}{Theorem}[section]

\newtheorem{lemma}[theorem]{Lemma}

\newtheorem{rmk}{Remark}[section]

\newcommand{\argmax}{\operatornamewithlimits{argmax}}

\DeclareMathOperator{\sgn}{sgn}

\numberwithin{equation}{section}

\begin{document}

\title{The Impact of Ambiguity on the Optimal Exercise Timing of Integral Option Contracts}

\author{Luis H. R. Alvarez E.\thanks{Department of Accounting and Finance, Turku School of Economics,
FIN-20014 University of Turku, Finland, E-mail: lhralv@utu.fi}\quad
S\"oren Christensen\thanks{Mathematisches Seminar,
Christian-Albrechts-Universit{\"a}t zu Kiel,
Ludewig-Meyn-Str. 4,
D-24098 Kiel,
Germany,  E-mail: christensen@math.uni-kiel.de}}
\maketitle

\abstract{We consider the impact of ambiguity on the optimal timing of a class of two-dimensional integral option contracts when the exercise payoff is a positively homogeneous measurable function. Hence, the considered class of exercise payoffs includes discontinuous functions as well. We identify a parameterized family of excessive functions generating an appropriate class of supermartingales for the considered problems and then express the value of the optimal policy as well as the worst case measure in terms of these processes. The advantage of our approach is that it reduces the analysis of the multidimensional problem to the analysis of an ordinary one-dimensional static optimization problem. In that way it simplifies earlier treatments of the problem without ambiguity considerably. We also illustrate our findings in explicitly parameterized examples. }\\



\thispagestyle{empty} \clearpage \setcounter{page}{1}

\section{Introduction}

Integral options play a crucial role in the valuation and optimal exercise timing of contracts written on cumulative reserves subject to stochastic growth. Put somewhat differently, if the growth rate of a reserve is a stochastic process itself, then determining the date at which the expected present value of the reserves are maximized constitutes an optimal stopping problem involving an integral option (cf. \cite{KrMo1994}). In this study our objective is to precisely focus on this problem in the presence of Knightian uncertainty when the growth rate of the underlying reserves follow a geometric Brownian motion.

The existing literature studying ambiguity and its impact on decision making is extensive ranging from studies based on atemporal multiple priors setting (cf. \citet{GiSch89}, \citet{Be02}, \citet{Kli_et_al_05},
\citet{Ma_et_al_06} and \citet{NiOz06}) to an intertemporal
recursive multiple priors setting  (cf. \cite{EpWa94}, \citet{ChEp02}, \cite{EpMi03}, and \citet{EpSch03}).
The impact of Knightian uncertainty on the optimal timing policies of ambiguity averse decision makers was first investigated by
\cite{NiOz04} in a job search model. This analysis has been subsequently generalized to various directions.
\cite{NiOz07} studied how Knightian uncertainty affects
optimal irreversible investment timing in a continuous time model based on geometric Brownian motion. \cite{Al07} extended the analysis in \cite{NiOz07} and analyzed the impact of Knightian uncertainty on monotone single boundary stopping problems and expressed the value as well as the optimality conditions for the stopping boundaries in terms of the monotone fundamental solutions generating the minimal excessive mappings for the ambiguous dynamics. \cite{Ri2009} developed
a general discrete time minmax martingale approach to optimal stopping problems in the presence of ambiguity aversion. These results were subsequently generalized to a continuous time setting in \cite{ChRi2013}, where the value is identified as the smallest right continuous $g$-martingale dominating the payoff process.
\cite{MiWa_11}, in turn, investigated how ambiguity affects optimal timing in a model based on a general discrete time Feller-continuous Markov process. A general analysis of the impact of ambiguity on timing when the underlying is a general regular diffusion was developed in \cite{Chr13}. In that study a parameterized class of minimal excessive mappings generating the worst case measure as well as the appropriate class of supermartingales is identified explicitly. \cite{Chr13} shows how the value and optimal timing policy can be expressed in terms of these mappings. \cite{EpJi2019} investigated optimal learning in the case where the underlying driving Brownian motion is subject to drift ambiguity and solved the optimal stopping problem characterizing the optimal learning policy explicitly. Finally, \cite{AlCh2019} analyzed the optimal stopping decisions of ambiguity averse decision makers in the case where the underlying is a two-dimensional geometric Brownian motion and the exercise payoff is a positively homogeneous and measurable function.

In this study we investigate the impact of Knightian uncertainty on the optimal timing decisions of an ambiguity averse decision maker when the exercise payoff is assumed to be a measurable and positively homogeneous function of the underlying reserve as well as its randomly fluctuating growth rate. Since this type of contracts are relatively common in practice, our results cast light on a large class of valuation problems. As both the underlying process itself and the integral capturing its cumulative value are involved, the underlying problem has first of all a two-dimensional structure. As the uncertainty of the decision maker directly just refers to the drift of the underlying asset price process, the underlying ambiguity structure is one-dimensional. It, however, affects both processes involved in the decision at the same time. Therefore, in most problems of interest, the identification of the worst case measure turns out to be a difficult task. To the best of our knowledge, such kind of structure has until now not been considered in the existing literature in detail.

Instead of tackling the considered stopping problem directly via standard dynamic programming arguments, we follow the analysis of \cite{AlCh2019} and consider the ratio of the underlying processes. Since this ratio constitutes a linear diffusion with known boundary behavior, we are in this way able to reduce the dimensionality of the considered problem due to the positive homogeneity of the exercise payoff. We then follow \cite{Chr13} and identify explicitly a class of excessive functions, parameterized by an arbitrary reference point, which can be utilized as numeraire assets and which generate the supermartingales needed for the determination of the optimal timing policy and its value as well as the worst case measure. In this way we show how the determination of the optimal policy and its value can be reduced into the analysis of the extremal points of a ratio depending on a single state variable and a reference point. Interestingly, we find that all the elements in the set of maximizers of the ratio are included into the stopping region of the considered problem. In this way our findings show how elements of the stopping region can be identified by simply determining the maximal points of a function depending on a single state variable. We also delineate circumstances under which the optimal policy is of a standard single boundary stopping type and under which it constitutes a two-boundary policy. We find that in both circumstances the worst case measure and optimal timing rule constitute a Nash equilibrium. In that way our findings show that the considered stopping problem of an ambiguity averse decision maker can be interpreted as a game between the decision maker and a malevolent opponent controlling the measures characterizing the probabilistic structure of the considered problem. In line with the findings of \cite{AlCh2019}, we find that Knightian uncertainty has a profound and nontrivial impact on the optimal timing policy. First of all, we find that ambiguity does not only affect the rate at which the underlying stochastic dynamics are evolving, it also impacts the rate at which an uncertainty averse decision maker discounts the exercise payoff under the worst case measure -- a phenomenon that does not arise in a one-dimensional setting. Second, in contrast with the findings of \cite{Chr13} focusing on linear diffusions, our results show that the state at which the density generator switches optimally from one extreme to another does not coincide with the reference point at which switching occurs in one-dimensional problems. In this way our findings show that the dimensionality of the problem strongly affects the nature of the solution. It is at this point worth emphasizing that since positively homogeneous functions are not necessarily continuous, our approach covers discontinuous payoffs as well. In this way we extend standard treatments of optimal stopping problems.

The considered stopping problem and the underlying stochastic dynamics are presented in Section 2. Our main results characterizing the value and optimal timing policy in different circumstances are the stated in Section 3. Our results are the explicitly illustrated in three different cases in Section 4. Finally, Section 5 concludes our study.

\section{Underlying Dynamics and Problem Setting}
Let $\mathbf{W}_{t}$ be an ordinary Brownian motion under the reference measure $\mathbb{P}$ and assume that the underlying process follow under the measure $\mathbb{P}$ the stochastic dynamics characterized by the stochastic differential equation
\begin{align}
dX_t &= \mu X_t dt + \sigma X_t dW_{t},\quad X_0=x\in \mathbb{R},
\end{align}
where $\mu \in \mathbb{R}$ and  $\sigma\in \mathbb{R}_{+}$ are known constants. Given the process $X_t$, we define the process $Y_t$ as
$$
Y_t=\int_0^t X_sds.
$$

Following standard approaches investigating the impact of Knightian uncertainty on optimal timing of contingent contracts, let the degree of ambiguity $\kappa>0$ be given and denote by $\mathcal{P}^\kappa$ the set of all probability measures, that are equivalent to $\mathbb{P}$ with density process of the form
$$
\mathcal{M}_t^{\theta}=e^{-\int_0^t \theta_s dW_{s} - \frac{1}{2}\int_0^t \theta_s^2 ds}
$$
for a progressively measurable process $\{\theta_t\}_{t\geq 0}$ satisfying the constraint $|\theta_{t}|\leq \kappa$ for all $t\geq 0$. Invoking the Cameron-Martin-Girsanov change of measure theorem show that under the measure $\mathbb{Q}^{\theta}$ defined by the likelihood ratio
$$
\frac{d\mathbb{Q}^{\theta}}{d\mathbb{P}}=\mathcal{M}_t^{\theta}
$$
we have that
\begin{align*}
\tilde{W}_{t}^{\theta} &= W_{t} + \int_0^t\theta_{s}ds
\end{align*}
is an ordinary $\mathbb{Q}^{\theta}$-Brownian motion. Thus, we notice that under a measure $\mathbb{Q}^{\theta}\in \mathcal{P}^\kappa$ the dynamics of the underlying process reads as
\begin{align}
dX_t = (\mu-\sigma\theta_{t}) X_t dt + \sigma X_t d\tilde{W}_{t}^{\theta},\quad X_0=x\in \mathbb{R}_+.\label{X}
\end{align}

Given the underlying processes and the class of equivalent measures generated by the density process $\mathcal{M}_t^{\theta}$, our objective is to now study the optimal stopping problem
\begin{align}
V_\kappa(x,y) = \sup_{\tau\in \mathcal{T}}\inf_{\mathbb{Q}^{\theta}\in \mathcal{P}^\kappa}\mathbb{E}_{\mathbf{x}}^{{\mathbb{Q}^{ \theta}}}\left[e^{-r\tau}F(X_\tau,Y_\tau)\mathbbm{1}_{\tau < \infty}\right],\label{stopping}
\end{align}
where $F:\mathbb{R}_+^2\mapsto \mathbb{R}$ is a known measurable function which is assumed to be positively homogeneous of degree one and $r>0$ is a known constant discount rate. Note that in the absence of ambiguity (i.e. when $\kappa=0$) and when $F(x,y)=y$ the model coincides with the integral option studied in \cite{KrMo1994} (see also \cite{MaYo2005a,MaYo2005b} and \cite{LeUr2007}).

\section{Analysis and Main Results}
The differential operator representing the generator of the process $(X,Y)$ reads under the measure $\mathbb{Q}^{\theta}\in \mathcal{P}^\kappa$ as
\begin{align}\label{generator}
\mathcal{A}^\theta=\frac{1}{2}\sigma^2x^2 \frac{\partial^2}{\partial x^2}+(\mu-\sigma\theta)x\frac{\partial}{\partial x}+x\frac{\partial}{\partial y}.
\end{align}
Assume now that $u:\mathbb{R}_+^2\mapsto \mathbb{R}_+$ is twice continuously differentiable on $\mathbb{R}_+^2$. We directly observe that
\begin{align*}
\left(\mathcal{A}^\theta u\right)(x,y)\geq \frac{1}{2}\sigma^2x^2 u_{xx}(x,y)+\mu x u_x(x,y)-\kappa\sigma  \sgn(u_x(x,y)) x u_x(x,y)+xu_y(x,y)
\end{align*}
for all admissible density generators $\theta$ and all $(x,y)\in \mathbb{R}_+^2$ and that
\begin{align*}
\left(\mathcal{A}^{\theta^\ast} u\right)(x,y)= \frac{1}{2}\sigma^2x^2 u_{xx}(x,y)+\mu x u_x(x,y)-\kappa\sigma  \sgn(u_x(x,y)) x u_x(x,y)+xu_y(x,y)
\end{align*}
for $\theta^\ast=\kappa\sgn(u_x(x,y))$.  Assume now that $u(x,y)$ satisfies the partial differential equation $(\mathcal{A}^{\theta^\ast} u)(x,y)=ru(x,y)$. Invoking the It{\^o}-D{\"o}blin theorem to $u(x,y)$ yields
\begin{align*}
e^{-rT}u(X_T,Y_T) &=u(x,y)+\int_0^T e^{-rs}\left(\kappa\sgn(u_x(X_s,Y_s))-\theta_s\right)\sigma X_su_x(X_s,Y_s)ds \\ &+\int_0^T e^{-rs}\sigma X_su_x(X_s,Y_s)dW_s^\theta\\
&\geq u(x,y)+\int_0^T e^{-rs}\sigma X_su_x(X_s,Y_s)dW_s^\theta
\end{align*}
with identity only when $\theta_t^\ast=\kappa\sgn(u_x(X_t,Y_t))$. Consequently, under $\mathbb{Q}^{\theta^\ast}$ $e^{-rt}u(X_t,Y_t)$ constitutes a positive local martingale.

Given the homogeneity of the exercise payoff, we make an ansatz that the value should be homogeneous as well. Let us, therefore, study solutions of the form $u(x,y)=x h(z), z=y/x,$ to the partial differential equation $(\mathcal{A}^{\theta^\ast} u)(x,y)=ru(x,y)$. We notice that now
$$
(\mathcal{A}^{\theta^\ast} u)(x,y)-ru(x,y)=x\left[\frac{1}{2}\sigma^2z^2h''(z) + (1-\mu z)h'(z)-(r-\mu)h(z)-\kappa\sigma\theta^\ast (h(z)-zh'(z))\right],
$$
where $\theta^\ast=\sgn(h(z)-zh'(z))$. It holds that
\begin{align}\label{lower}
\frac{1}{2}\sigma^2z^2h''(z) + (1-\mu z+\kappa\sigma z)h'(z)-(r-\mu+\kappa\sigma)h(z)=0
\end{align}
on the set $\{z\in \mathbb{R}_+:h(z)>zh'(z)\}$ and
\begin{align}\label{upper}
\frac{1}{2}\sigma^2z^2h''(z) + (1-\mu z-\kappa\sigma z)h'(z)-(r-\mu-\kappa\sigma)h(z)=0
\end{align}
on the set $\{z\in \mathbb{R}_+:h(z)<zh'(z)\}$.\\

\begin{rmk} It is clear that the differential equations \eqref{lower} and \eqref{upper} are associated with diffusions of the type
$$
dZ_t=(1 - \delta_i Z_t)dt + \sigma Z_t dW_t,\quad i=1,2,
$$
where $\delta_1 = \mu-\kappa\sigma$ and $\delta_2 = \mu+\kappa\sigma$. Defining the process $L_t=Z_t^{-1}$ and applying the It\^o-D{\"o}blin theorem
shows that
\begin{align}\label{logistic}
dL_t=(\sigma^2+\delta_i-L_t)L_tdt-\sigma L_t dW_t
\end{align}
illustrating how the controlled dynamics are associated with the logistic diffusion characterized by \eqref{logistic}.
\end{rmk}

Let
\begin{align*}
\psi_{\kappa}&=-\frac{1}{2}-\frac{\mu-\kappa\sigma}{\sigma^2}+\sqrt{\left(\frac{1}{2}+\frac{\mu-\kappa\sigma}{\sigma^2}\right)^2+
\frac{2(r-\mu+\kappa\sigma)}{\sigma^2}}\\
\varphi_{\kappa}&=-\frac{1}{2}-\frac{\mu-\kappa\sigma}{\sigma^2}-\sqrt{\left(\frac{1}{2}+\frac{\mu-\kappa\sigma}{\sigma^2}\right)^2+
\frac{2(r-\mu+\kappa\sigma)}{\sigma^2}}
\end{align*}
denote the roots of the quadratic equation
$$
q^2 + \left(1+\frac{2(\mu-\kappa\sigma)}{\sigma^2}\right)q-\frac{2(r-\mu+\kappa\sigma)}{\sigma^2}=0
$$
and define the functions
\begin{align*}
P_{\kappa}(z)&=\left(\frac{2}{\sigma^2 z}\right)^{\psi_\kappa}U\left(\psi_{\kappa}, 1+\psi_{\kappa}-\varphi_{\kappa},\frac{2}{\sigma^2 z}\right)\\
Q_{\kappa}(z)&=\left(\frac{2}{\sigma^2 z}\right)^{\psi_\kappa}M\left(\psi_{\kappa}, 1+\psi_{\kappa}-\varphi_{\kappa},\frac{2}{\sigma^2 z}\right),
\end{align*}
where $M$ denotes {\em Kummer's} confluent hypergeometric and $U$ denotes {\em Tricomi's} confluent hypergeometric function.
It is now a straightforward, yet somewhat laborious, exercise in solving ODE's to show that utilizing the transformation $h(z)=(2/(\sigma^2 z))^qf(2/(\sigma^2 z))$ results in
\begin{align*}
h(z) = c_1 P_\kappa(z)+c_2Q_\kappa(z)
\end{align*}
on the set $\{z\in \mathbb{R}_+:h(z)>zh'(z)\}$ and
\begin{align*}
h(z) = c_1 P_{-\kappa}(z)+c_2Q_{-\kappa}(z)
\end{align*}
on the set $\{z\in \mathbb{R}_+:h(z)<zh'(z)\}$. Before proceeding in the analysis of the function $h$ we first establish the strict convexity of the fundamental solutions and characterize the
behavior of the elasticity of $P_{\kappa}(z)$ on $\mathbb{R}_+$.
\begin{lemma}\label{l1}
\begin{itemize}
  \item[(A)] The fundamental solutions $P_\kappa(z),Q_\kappa(z),P_{-\kappa}(z),Q_{-\kappa}(z)$ are strictly convex on $\mathbb{R}_+$.
  \item[(B)] Equation $P_{\kappa}(z)-zP_{\kappa}'(z)=0$ has a unique root $\bar{z}_\kappa>1/r$ and $\bar{z}_\kappa=\argmax\{z/P_\kappa(z)\}$.
\end{itemize}
\end{lemma}
\begin{proof}
(A) We first observe that the lower boundary $0$ is entrance and the upper boundary is natural for the underlying diffusion process
$$
dZ_t=(1 - (\mu+\kappa\sigma) Z_t)dt + \sigma Z_t dW_t,\quad Z_0=z.
$$
Reordering now the ordinary differential equation \eqref{lower} shows that for $P_{\kappa}(z)$ it holds
\begin{align}\label{modupper}
\frac{1}{2}\sigma^2 z^2 \frac{P_{\kappa}''(z)}{S_\kappa'(z)}=\rho_\kappa \frac{P_{\kappa}(z)-zP_{\kappa}'(z)}{S_\kappa'(z)}-(1-rz)\frac{P_{\kappa}'(z)}{S_\kappa'(z)}
\end{align}
where $\rho_k=r-\mu +\kappa\sigma$ and
$$
S_\kappa'(z)=z^{\frac{2(\mu-\kappa\sigma)}{\sigma^2}}e^{\frac{2}{\sigma^2 z}}.
$$
Since
\begin{align}\label{canonical}
\frac{d}{dz}\frac{P_{\kappa}'(z)}{S_\kappa'(z)}=\rho_\kappa P_{\kappa}(z)m_\kappa'(z),
\end{align}
where $m_\kappa'(z)=2/(\sigma^2z^2 S_\kappa'(z))$, we find that
\begin{align*}
\frac{P_{\kappa}'(z)}{S_\kappa'(z)}-\frac{P_{\kappa}'(a)}{S_\kappa'(a)}=\int_a^z\rho_\kappa P_{\kappa}(y)m_\kappa'(y)dy,
\end{align*}
where $z>a>0$.
On the other hand, since
\begin{align*}
\frac{d}{dz}\frac{P_{\kappa}(z)-zP_{\kappa}'(z)}{S_\kappa'(z)}=(1-rz)P_{\kappa}(z)m_\kappa'(z)
\end{align*}
we find  that
\begin{align*}
\rho_\kappa\frac{P_{\kappa}(z)-zP_{\kappa}'(z)}{S_\kappa'(z)}-\rho_\kappa\frac{P_{\kappa}(a)-aP_{\kappa}'(a)}{S_\kappa'(a)}=\rho_\kappa \int_a^z (1-ry)P_{\kappa}(y)m_\kappa'(y)dy.
\end{align*}
Plugging these results into \eqref{modupper} and simplifying yields
\begin{align*}
\frac{1}{2}\sigma^2 z^2 \frac{P_{\kappa}''(z)}{S_\kappa'(z)}&= \rho_\kappa\frac{P_{\kappa}(a)-aP_{\kappa}'(a)}{S_\kappa'(a)}-(1-rz)\frac{P_{\kappa}'(a)}{S_\kappa'(a)}+
\rho_\kappa\int_a^z r(z-y)P_{\kappa}(y)m_\kappa'(y)dy\\
&>\rho_\kappa\frac{P_{\kappa}(a)-aP_{\kappa}'(a)}{S_\kappa'(a)}-(1-rz)\frac{P_{\kappa}'(a)}{S_\kappa'(a)}.
\end{align*}
Since $\lim_{a\rightarrow 0+}P_{\kappa}'(a)/S_\kappa'(a)=0,\lim_{a\rightarrow 0+}P_{\kappa}(a)=1$, and $\lim_{a\rightarrow 0+}S_{\kappa}'(a)=\infty$ we find by letting $a\downarrow 0$
that $P_{\kappa}''(z)>0$ for all $z>0$ proving the alleged strict convexity of $P_{\kappa}(z)$. Establishing the strict convexity of $P_{-\kappa}(z),Q_{\kappa}(z)$, and $Q_{-\kappa}(z)$
is completely analogous.\\

(B) It is clear from our analysis above that
$$
\frac{P_{\kappa}(z)-zP_{\kappa}'(z)}{S_\kappa'(z)}=\int_0^z(1-ry)P_{\kappa}(y)m_\kappa'(y)dy
$$
showing that $P_{\kappa}(z)-zP_{\kappa}'(z)>0$ for all $z\leq 1/r$. Assume now that $z>x>1/r$. We then notice by utilizing the monotonicity of $1-rz$ and the canonical form  \eqref{canonical} that
\begin{align*}
\frac{P_{\kappa}(z)-zP_{\kappa}'(z)}{S_\kappa'(z)}&=\frac{P_{\kappa}(x)-xP_{\kappa}'(x)}{S_\kappa'(x)}+\int_x^z(1-ry)P_{\kappa}(y)m_\kappa'(y)dy\\
&\leq \frac{P_{\kappa}(x)-xP_{\kappa}'(x)}{S_\kappa'(x)}+\frac{(1-rx)}{\rho_\kappa}\left(\frac{P_{\kappa}'(z)}{S_\kappa'(z)}-\frac{P_{\kappa}'(x)}{S_\kappa'(x)}\right)\downarrow -\infty,
\end{align*}
as $z\rightarrow \infty$. The monotonicity and continuity of $(P_{\kappa}(z)-zP_{\kappa}'(z))/S_\kappa'(z)$ on $(1/r,\infty)$ now proves the alleged existence and uniqueness
of a root $\bar{z}_\kappa>1/r$. Since
$$
\frac{d}{dz}\left(\frac{z}{P_\kappa(z)}\right)=\frac{P_\kappa(z)-zP_\kappa'(z)}{P_{\kappa}^2(z)}
$$
we notice that $\bar{z}_\kappa$ is the unique maximizer of the ratio $z/P_\kappa(z)$.
\end{proof}

Given our observations above, we now follow \cite{AlCh2019} and let $c\in [0,\infty)$ be an arbitrary reference point. Define the function
$U_{c}:\mathbb{R}_+\mapsto \mathbb{R}_+$ as
\begin{align}\label{minimal}
U_{c}(z)=\begin{cases}
c_1(\hat{z}_{c}) P_{-\kappa}(z)+c_2(\hat{z}_{c}) Q_{-\kappa}(z),&z\geq\hat{z}_c,\\
B_{\kappa}^{-1}\left[\frac{P_{\kappa}'(c)}{S_\kappa'(c)}Q_\kappa(z)-\frac{Q_{\kappa}'(c)}{S_\kappa'(c)}P_\kappa(z)\right],&z\leq \hat{z}_{c},
\end{cases}
\end{align}
where $\hat{z}_{c}>c$ satisfies the condition
\begin{align}\label{switch}
\frac{P_{\kappa}'(c)}{S_\kappa'(c)}(Q_\kappa'(\hat{z}_{c})\hat{z}_{c}-Q_\kappa(\hat{z}_{c}))=
\frac{Q_{\kappa}'(c)}{S_\kappa'(c)}(P_\kappa'(\hat{z}_{c})\hat{z}_{c}-P_\kappa(\hat{z}_{c})),
\end{align}
$$
B_{\kappa} = \frac{\Gamma(\psi_\kappa-\varphi_\kappa+1)}{\Gamma(\psi_\kappa)}\left(\frac{2}{\sigma^2}\right)^{\psi_\kappa+\varphi_\kappa}>0
$$
denotes the constant Wronskian of the fundamental solutions with respect to the scale $S_\kappa'(z)$,
and
\begin{align*}
c_1(\hat{z}_{c}) &= \left(\frac{Q_{-\kappa}(\hat{z}_{c})-Q_{-\kappa}'(\hat{z}_{c})\hat{z}_{c}}{B_{-\kappa}S_{-\kappa}'(\hat{z}_{c})}\right)U_{c}'(\hat{z}_{c}-),\\
c_2(\hat{z}_{c}) &= \left(\frac{P_{-\kappa}'(\hat{z}_{c})\hat{z}_{c}-P_{-\kappa}(\hat{z}_{c})}{B_{-\kappa}S_{-\kappa}'(\hat{z}_{c})}\right)U_{c}'(\hat{z}_{c}-).
\end{align*}
Before proceeding in our analysis it is worth shortly explaining how the function $U_c$ is constructed. On $(0,\hat{z}_{c})$ the function $U_c(z)$ constitutes the solution of the
ordinary differential equation \eqref{lower} subject to the boundary conditions $U_c(c)=1$ and $U_c'(c)=0$. On $(\hat{z}_{c},\infty)$ the function $U_c(z)$, in turn, constitutes the solution of the
ordinary differential equation \eqref{upper} subject to the smoothness conditions $U_c(\hat{z}_{c}-)=U_c(\hat{z}_{c}+)$ and $U_c'(\hat{z}_{c}-)=U_c'(\hat{z}_{c}+)$ across the boundary $\hat{z}_{c}$.
Finally, the boundary $\hat{z}_c$ is determined from the condition $U_c'(\hat{z}_{c})\hat{z}_{c}=U_c(\hat{z}_{c})$. We are now in position to show the following auxiliary result characterizing
the smoothness and convexity properties of $U_c$.

\begin{lemma}\label{auxconv}
For each reference point $c\in[0,\infty)$, Equation \eqref{switch} has a unique root $\hat{z}_{c}>c$. Moreover, the function $U_c:\mathbb{R}_+\mapsto \mathbb{R}_+$ is twice
continuously differentiable and strictly convex.
\end{lemma}
\begin{proof}
We first establish that equation \eqref{switch} has a unique root $\hat{z}_{c}>c$ for all reference points $c\in[0,\infty)$. To see that this is indeed the case, consider the
behavior of the continuously differentiable function
\begin{align*}
D_{c}(z)=U_c'(z)z-U_c(z)=B_{\kappa}^{-1}\left[\frac{P_{\kappa}'(c)}{S_\kappa'(c)}\left(Q_\kappa'(z)z-Q_\kappa(z)\right)-
\frac{Q_{\kappa}'(c)}{S_\kappa'(c)}\left(P_\kappa'(z)z-P_\kappa(z)\right)\right].
\end{align*}
It is clear that $D_c(c)=-1<0$ and
$$
D_c'(z)=B_{\kappa}^{-1}\left[\frac{P_{\kappa}'(c)}{S_\kappa'(c)}Q_\kappa''(z)z-
\frac{Q_{\kappa}'(c)}{S_\kappa'(c)}P_\kappa''(z)z\right]>0
$$ for all $z\geq c$. Part (B) of Lemma \ref{l1}, in turn, implies that
$(P_\kappa'(z)z-P_\kappa(z))/S_\kappa'(z)\rightarrow \infty$ as $z\rightarrow\infty$. Noticing now that
$$
\frac{d}{dz}\left[\frac{Q_\kappa'(z)z-Q_\kappa(z)}{S_\kappa'(z)}\right]=-(1-rz)Q_{\kappa}(z)m_\kappa'(z) \gtreqqless 0,\quad z \gtreqqless \frac{1}{r}
$$
demonstrates that
$$
\frac{Q_\kappa'(1/r)/r-Q_\kappa(1/r)}{S_\kappa'(1/r)}<\frac{Q_\kappa'(z)z-Q_\kappa(z)}{S_\kappa'(z)}<0.
$$
Consequently, we notice that $\lim_{z\rightarrow\infty}D_{c}(z)=\infty$ proving the existence of $\hat{z}_\kappa$. The alleged uniqueness of  $\hat{z}_\kappa$ follows from the monotonicity of $D_c(z)$.
To prove the alleged smoothness of the function $U_c$, we first notice that the function $U_c(z)$ is by construction continuously differentiable on $\mathbb{R}_+$ and twice
continuously differentiable on $\mathbb{R}_+\backslash\{\hat{z}_c\}$. However, since
\begin{align*}
\frac{1}{2}\sigma^2 \hat{z}_c^2U_c''(\hat{z}_c-)=rU_c(\hat{z}_c-)-U_c'(\hat{z}_c-)=rU_c(\hat{z}_c+)-U_c'(\hat{z}_c+)=\frac{1}{2}\sigma^2 \hat{z}_c^2U_c''(\hat{z}_c+)
\end{align*}
we find that $U_c(z)$ is twice continuously differentiable on the entire $\mathbb{R}_+$. Finally, since $Q_\kappa'(c)<0<P_\kappa'(c)$ we notice that the convexity of the fundamental
solutions $P_\kappa(z)$ and $Q_\kappa(z)$ guarantee that $U_c(z)$ is strictly convex
on $(0,\hat{z}_c]$. On the other hand, the twice continuous differentiability of $U_c(z)$, the strict convexity of $U_c(z)$ on $(0,\hat{z}_c]$, and the condition
$U_c'(\hat{z}_c)\hat{z}_c=U_c(\hat{z}_c)$ guarantee that
$c_2(\hat{z}_c)>0$ proving the strict convexity of $U_c(z)$ on $(\hat{z}_c,\infty)$ as well.
\end{proof}

\begin{rmk}
It is worth noticing that Part (B) of Lemma \ref{l1} implies that in Lemma \ref{auxconv} we have $\lim_{c\rightarrow 0+}\hat{z}_c = \bar{z}_\kappa$. Consequently,
\begin{align*}
U_{0}(z)=\begin{cases}
c_1(\bar{z}_\kappa) P_{-\kappa}(z)+c_2(\bar{z}_\kappa) Q_{-\kappa}(z),&z\geq\bar{z}_\kappa,\\
P_\kappa(z),&z\leq \bar{z}_\kappa.
\end{cases}
\end{align*}
This observation follows from the fact that $0$ is entrance for the underlying ratio-process $Z_t=Y_t/X_t$ and $\lim_{z\rightarrow 0}P_\kappa(z)=1$.
\end{rmk}

\begin{lemma}\label{supermartingale}
Denote by $\mathbb{Q}^{\theta^c}\in \mathcal{P}^\kappa$ the measure induced by the density generator $$\theta_t^c:=\kappa\sgn(\hat{z}_c X_t-Y_t),$$ where $c\in[0,\infty)$ and $\hat{z}_c$
denotes the unique root of \eqref{switch}. Then,
\begin{align}
dX_t&=(\mu-\kappa\sigma\sgn(\hat{z}_c X_t-Y_t))X_tdt+\sigma X_t d\tilde{W}_t^{\theta^c},\label{sde1}
\end{align}
and
\begin{align}\label{ratioprocess}
\begin{split}
dZ_t &=(1-(\mu-\sigma^2-\kappa\sigma\sgn(\hat{z}_c -Z_t))Z_t)dt-\sigma Z_t d\tilde{W}_{t}^{\theta^c},
\end{split}
\end{align}
where $\tilde{W}_{t}^{\theta^c}$ is a standard Brownian motion under the measure $\mathbb{Q}^{\theta^c}$. Moreover, for any stopping time
$\tau\in \mathcal{T}$, admissible density generator $\theta$, and $(x,y)\in \mathbb{R}_+^2$ we have
$$
\mathbb{E}_{\mathbf{x}}^{\mathbb{Q}^{\theta^c}}\left[e^{-r\tau}X_\tau U_{c}(Y_\tau/X_\tau)\mathbbm{1}_{\{\tau<\infty\}}\right]\leq x U_c(y/x)\leq
\mathbb{E}^{\mathbb{Q}^{\theta}}_{\mathbf{x}}\left[e^{-r\tau}X_\tau U_{c}(Y_\tau/X_\tau)\mathbbm{1}_{\{\tau<\infty\}}\right].
$$
\end{lemma}
\begin{proof}
Applying the It{\^o}-D{\"o}blin theorem to the mapping $(t,x,y)\mapsto e^{-rt}xU_c(y/x)$ yields
\begin{align*}
e^{-rt}X_{t}U_c(Z_t)=xU_c(z)+\int_0^{t}e^{-rs}\sigma X_s\Delta_c(Z_s)\left(\kappa\sgn(\hat{z}_c-Z_s)-\theta_s\right)ds+\int_0^{t}e^{-rs}\sigma X_s\Delta_c(Z_s)d\tilde{W}_t^{\theta},
\end{align*}
where $\Delta_c(z)=U_c(z)-U_c'(z)z$ and $Z_t=Y_t/X_t$. Since $\Delta_c(Z_t)(\kappa\sgn(\hat{z}_c-Z_t)-\theta_t)\geq 0$ for all admissible density generators $\theta_t$ and all $t\geq 0$, we notice that
\begin{align*}
e^{-rt}X_{t}U_c(Z_t)\geq xU_c(z)+\int_0^{t}e^{-rs}\sigma X_s\Delta_c(Z_s)d\tilde{W}_t^{\theta}.
\end{align*}
Assume that $G\subset\mathbb{R}_+^2$ is an open subset with compact closure in $\mathbb{R}_+^2$ and let
$\tau_G=\inf\{t\geq0: (X_t,Y_t)\not\in G\}$ denote the first exit time of the process $(X_t,Y_t)$ from $G$. We notice that the stopped process
$\{e^{-r(t\wedge \tau_G)}X_{t\wedge \tau_G}U_c(Z_{t\wedge \tau_G})\}_{t\geq 0}$ is a bounded positive $\mathbb{Q}^{\theta}$-submartingale.
For $\theta_t^c = \kappa \sgn(\hat{z}_c-Z_t)$ we have
\begin{align*}
e^{-rt}X_{t}U_c(Z_t)= xU_c(z)+\int_0^{t}e^{-rs}\sigma X_s\Delta_c(Z_s)d\tilde{W}_t^{\theta^c}
\end{align*}
proving that the process $\{e^{-r(t\wedge \tau_G)}X_{t\wedge \tau_G}U_c(Z_{t\wedge \tau_G})\}_{t\geq 0}$ is a bounded positive local $\mathbb{Q}^{\theta^c}$-martingale. Analogous computations demonstrate that the process $\{e^{-rt}X_{t}U_c(Z_{t})\}_{t\geq 0}$ is actually a positive $\mathbb{Q}^{\theta^c}$-martingale and, therefore, a supermartingale. Finally, it is clear that under the measure $\mathbb{Q}^{\theta^c}$ the underlying process $X$ as well as the ratio $Z$ evolve according to the random dynamics characterized by the stochastic differential equations \eqref{sde1}, and \eqref{ratioprocess}.
\end{proof}

\begin{lemma}\label{decreasingsupermartingale}
Denote by $\mathbb{Q}^{\kappa}\in \mathcal{P}^\kappa$ the measure induced by the density generator $\theta_t=\kappa$ for all $t\geq 0$ and let $U_\infty(z)=Q_\kappa(z)$. Then,
\begin{align}
dX_t=(\mu-\kappa\sigma)X_tdt+\sigma X_t d\tilde{W}_t^{\kappa},\label{sde1mod}
\end{align}
and
\begin{align}\label{ratioprocessmod}
dZ_t =(1-(\mu-\sigma^2-\kappa\sigma)Z_t)dt-\sigma Z_t d\tilde{W}_{t}^{\kappa},
\end{align}
where $\tilde{W}_{t}^{\kappa}$ is a standard Brownian motion under the measure $\mathbb{Q}^{\kappa}$. Moreover, for any stopping time
$\tau\in \mathcal{T}$, admissible density generator $\theta$, and $(x,y)\in \mathbb{R}_+^2$ we have
$$
\mathbb{E}_{\mathbf{x}}^{\mathbb{Q}^{\kappa}}\left[e^{-r\tau}X_\tau U_{\infty}(Y_\tau/X_\tau)\mathbbm{1}_{\{\tau<\infty\}}\right]\leq x U_\infty(y/x)\leq
\mathbb{E}^{\mathbb{Q}^{\theta}}_{\mathbf{x}}\left[e^{-r\tau}X_\tau U_{\infty}(Y_\tau/X_\tau)\mathbbm{1}_{\{\tau<\infty\}}\right].
$$
\end{lemma}
\begin{proof}
The stochastic differential equations \eqref{sde1mod} and \eqref{ratioprocessmod} follow directly from the definition of the processes by setting $\theta_t=\kappa$ for all $t\geq 0$. Applying the It{\^o}-D{\"o}blin theorem to the mapping $(t,x,y)\mapsto e^{-rt}xQ_\kappa(y/x)$ yields
\begin{align*}
e^{-rt}X_{t}Q_\kappa(Z_t)=xQ_\kappa(z)+\int_0^{t}e^{-rs}\sigma X_s\Delta(Z_s)\left(\kappa-\theta_s\right)ds+\int_0^{t}e^{-rs}\sigma X_s\Delta(Z_s)d\tilde{W}_t^{\theta},
\end{align*}
where $\Delta(z)=Q_\kappa(z)-Q_\kappa'(z)z > 0$ due to the monotonicity and positivity of $Q_\kappa(z)$. Hence,
\begin{align*}
e^{-rt}X_{t}Q_\kappa(Z_t)\geq xQ_\kappa(z)+\int_0^{t}e^{-rs}\sigma X_s\Delta(Z_s)d\tilde{W}_t^{\theta}
\end{align*}
for any admissible density generator $\theta_t$ with identity
\begin{align*}
e^{-rt}X_{t}Q_\kappa(Z_t)= xQ_\kappa(z)+\int_0^{t}e^{-rs}\sigma X_s\Delta(Z_s)d\tilde{W}_t^{\kappa}
\end{align*}
only when $\theta_t=\kappa$ for all $t\geq 0$. Establishing now the alleged inequalities is identical with the proof of Lemma \ref{supermartingale}.
\end{proof}

Lemma \ref{supermartingale} and Lemma \ref{decreasingsupermartingale} characterize the appropriate worst case supermartingales for the considered class of problems in the present setting. It is worth pointing out that in strong contrast to the one dimensional diffusion case studied in \cite{Chr13}, the minimum point $c$ of the function $U_c$ and the switching point of the density generator $\theta^c$ do not coincide in the present case.

 As in \cite{AlCh2019} these can be efficiently utilized in the characterization of the stopping set as well as the value of the optimal policy under the worst case measure. To see that this is indeed the case, we now define for all $c\in \mathbb{R}_+$ and $z\in \mathbb{R}_+$ the function
$$
\Pi_c(z)=\frac{F(1,z)}{U_c(z)}.
$$
Our first characterization extending the finding of Proposition 3.3 in \cite{AlCh2019} to the integral option setting is proved in the following.
\begin{theorem}\label{stoppingset}
 If
$$
z_{c}^\ast\in \argmax\left\{\Pi_c(z)\right\},
$$
then $\{(x,y)\in \mathbb{R}_+^2: y=z_{c}^\ast x\}\subseteq\Gamma_\kappa:=\{(x,y)\in \mathbb{R}_+^2: V_\kappa(x,y)=F(x,y)\}$.
\end{theorem}
\begin{proof}
	Assume that the set $\argmax\left\{\Pi_c(z)\right\}\neq \emptyset$ and let
	$
	z_{c}^\ast\in \argmax\left\{\Pi_c(z)\right\}.
	$
	Relying now on part (i) of Theorem 3 in \cite{Chr13} shows
	\begin{align*}
	\inf_{\mathbb{Q}^{\theta}\in \mathcal{P}^\kappa}\mathbb{E}_{\mathbf{x}}^{{\mathbb{Q}^{\theta}}}\left[e^{-r\tau}F(X_\tau,Y_\tau)\mathbbm{1}_{\{\tau<\infty\}}\right]&=
\inf_{\mathbb{Q}^{\theta}\in \mathcal{P}^\kappa}\mathbb{E}_{\mathbf{x}}^{{\mathbb{Q}^{\theta}}}\left[e^{-r\tau}X_\tau U_c(Z_\tau)\Pi_c(Z_\tau)\mathbbm{1}_{\{\tau<\infty\}}\right]\\
	&\leq \Pi_c(z_{c}^\ast)\inf_{\mathbb{Q}^{\theta}\in \mathcal{P}^\kappa}\mathbb{E}_{\mathbf{x}}^{{\mathbb{Q}^{\theta}}}\left[e^{-r\tau}X_\tau U_c(Z_\tau)\mathbbm{1}_{\{\tau<\infty\}}\right]\\
	&= \Pi_c(z_{c}^\ast)\mathbb{E}_{\mathbf{x}}^{{\mathbb{Q}^{\theta^c}}}\left[e^{-r\tau}X_\tau U_c(Z_\tau)\mathbbm{1}_{\{\tau<\infty\}}\right]\\
	&\leq \Pi_c(z_{c}^\ast)xU_c(z)
	\end{align*}
	for all $\tau\in \mathcal{T}$, $(x,y)\in\mathbb{R}_+^2$, and $c\in \mathbb{R}_+$. Hence, we have established that $V_\kappa(x,y) \leq \Pi_c(z^\ast)xU_c(z)$. On the other hand, since $V_\kappa(x,y) \geq F(x,y)$ for all $(x,y)\in \mathbb{R}_+^2$ we find
	$$
	\Pi_c(z)xU_c(z)\leq V_\kappa(x,y) \leq \Pi_c(z^\ast)xU_c(z)
	$$
	proving that $\{(x,y)\in \mathbb{R}_+^2: y=z_{c}^\ast x\}\subseteq\Gamma_\kappa:=\{(x,y)\in \mathbb{R}_+^2: V_\kappa(x,y)=F(x,y)\}$ as claimed.
\end{proof}
Theorem \ref{stoppingset} demonstrates how elements in the stopping set can be identified by analyzing the extremal points of the ratio $\Pi_c(z)$.
As intuitively is clear, this characterization can be utilized under slightly stronger conditions to characterize the value on subsets of the continuation region. This is proved in our next theorem extending the findings of Theorem 3.5 in \cite{AlCh2019} to the present case.
\begin{theorem}[Two-sided stopping]\label{t1}
	Assume that there are two points $z_{ic}^\ast\in \argmax\left\{\Pi_{c^\ast}(z)\right\}$, $i=1,2$, for some $c^\ast\in \mathbb{R}_+$ such that $\Pi_{c^\ast}(z_{1c}^\ast)=\Pi_{c^\ast}(z_{2c}^\ast)$. Then
\begin{itemize}
  \item[(A)]
$V_\kappa(x,y) =
\Pi_{c^\ast}(z_{ic}^\ast)xU_{c^\ast}(y/x) \mbox{ for all $(x,y)\in\mathbb{R}_+^2$ with }y/x\in (z_{1c}^\ast,z_{2c}^\ast)\mbox{ and $i=1,2$}.
$
  \item[(B)] If $\Pi_{c^\ast}(z_{ic}^\ast)>\Pi_{c^\ast}(z)$ for all $z\in (z_{1c}^\ast,z_{2c}^\ast)$ and $i=1,2$, then $$\{(x,y)\in \mathbb{R}_+^2: z_{1c}^\ast x < y < z_{2c}^\ast x\}\subseteq C_\kappa:=\{(x,y)\in \mathbb{R}_+^2:V_\kappa(x,y)>F(x,y)\}.$$
  \item[(C)] Let $\tau^\ast\in \mathcal T$ be such that  $\tau^\ast=\inf\{t\geq 0:Z_t\not\in (z_{1c}^\ast,z_{2c}^\ast)\}$ $\mathbb P_{\bf x}$-a.s. for all initial points $(x,y)$ with $z_{1c}^\ast x < y < z_{2c}^\ast x$. Then, $(\mathbb{Q}^{{\bm \theta}^{c^\ast}},\tau^\ast)$ is a Nash equilibrium in the sense that for all initial points $(x,y)$ with $z_{1c}^\ast x < y < z_{2c}^\ast x$ it holds that
  \begin{align*}
  \mathbb{E}_{\mathbf{x}}^{{\mathbb{Q}^{\theta}}}\left[e^{-r\tau^{\ast}}F(X_{\tau^{\ast}},Y_{\tau^{\ast}})\mathbbm{1}_{\{\tau^{\ast}<\infty\}}\right]&\geq \mathbb{E}_{\mathbf{x}}^{\mathbb{Q}^{\theta^{c^\ast}}}\left[e^{-r\tau^{\ast}}F(X_{\tau^{\ast}},Y_{\tau^{\ast}})\mathbbm{1}_{\{\tau^{\ast}<\infty\}}\right]\mbox{ for all }\mathbb{Q}^{\theta}\in \mathcal{P}^\kappa\\
  \mathbb{E}_{\mathbf{x}}^{\mathbb{Q}^{\theta^{c^\ast}}}\left[e^{-r\tau}F(X_{\tau},Y_{\tau})\mathbbm{1}_{\{\tau<\infty\}}\right]&\leq \mathbb{E}_{\mathbf{x}}^{\mathbb{Q}^{\theta^{c^\ast}}}\left[e^{-r\tau^{\ast}}F(X_{\tau^{\ast}},Y_{\tau^{\ast}})\mathbbm{1}_{\{\tau^{\ast}<\infty\}}\right]\mbox{ for all }\tau\in\mathcal T.
  \end{align*}
\end{itemize}
\end{theorem}
\begin{proof}
	As established in Lemma \ref{supermartingale}, the process
	$$
	M_t=e^{-rt}X_tU_{c^\ast}(Y_t/X_t)
	$$
	is a positive $\mathbb{Q}^{\theta^{c^\ast}}$-martingale. Consequently, for all $\tau\in\mathcal T$ and all initial points $(x,y)$ with $z_{1c}^\ast x < y < z_{2c}^\ast x$
	\begin{align*} \mathbb{E}_{\mathbf{x}}^{\mathbb{Q}^{\theta^{c^\ast}}}\left[e^{-r\tau}F(X_{\tau},Y_{\tau})\mathbbm{1}_{\{\tau<\infty\}}\right]&=\mathbb{E}_{\mathbf{x}}^{\mathbb{Q}^{\theta^{c^\ast}}}\left[M_\tau \Pi_{c^\ast}(Z_\tau)\mathbbm{1}_{\{\tau<\infty\}}\right]\\
	&\leq \sup_z \Pi_{c^\ast}(z)\mathbb{E}_{\mathbf{x}}^{\mathbb{Q}^{\theta^{c^\ast}}}\left[M_\tau \mathbbm{1}_{\{\tau<\infty\}}\right]\\
	&\leq\sup_z \Pi_{c^\ast}(z)\mathbb{E}_{\mathbf{x}}^{\mathbb{Q}^{\theta^{c^\ast}}}\left[M_0\right]\\
	&=\Pi_{c^\ast}(z_{ic}^\ast)xU_{c^\ast}(y/x).
	\end{align*}
Since the drift and diffusion coefficient of the controlled dynamics have no singularities on the interior of $\mathbb{R}_+$ it furthermore holds that $\tau^\ast$ as given in (C) is $\mathbb{Q}^{\theta^{c^\ast}}$-a.s. finite for all initial points $(x,y)$ with $z_{1c}^\ast x < y < z_{2c}^\ast x$
	and hence
	 $Z_{\tau^*}\in\{z_1,z_2\}$. Optional sampling now yields
	 \[\mathbb{E}_{\mathbf{x}}^{\mathbb{Q}^{\theta^{c^\ast}}}\left[M_{\tau^\ast} \mathbbm{1}_{\{\tau^\ast<\infty\}}\right]=\mathbb{E}_{\mathbf{x}}^{\mathbb{Q}^{\theta^{c^\ast}}}\left[M_0\right].\]
	 Therefore, we notice that for $\tau=\tau^*$ both inequalities in the calculations above are indeed equations, i.e.
	 \begin{align*}
	\mathbb{E}_{\mathbf{x}}^{\mathbb{Q}^{\theta^{c^\ast}}}\left[e^{-r\tau^\ast}F(X_{\tau^\ast},Y_{\tau^\ast})\mathbbm{1}_{\{\tau^\ast<\infty\}}\right]
	 &=\Pi_{c^\ast}(z_i)xU_{c^\ast}(y/x).
	 \end{align*}
	 	 Since $\mathbb{Q}^{\theta^{c^\ast}}\in \mathcal{P}^\kappa$ we notice that
	  $$
	 \inf_{\mathbb{Q}^{\theta}\in \mathcal{P}^\kappa}\mathbb{E}_{\mathbf{x}}^{\mathbb{Q}^{\theta}}\left[e^{-r\tau}F(X_\tau,Y_\tau)\right]
	 \leq \mathbb{E}_{\mathbf{x}}^{\mathbb{Q}^{\theta^{c^\ast}}}\left[e^{-r\tau}F(X_\tau,Y_\tau)\right]
	 $$
	 implying that $V_\kappa(x,y)\leq \Pi_{c^\ast}(z_{ic}^\ast)xU_{c^\ast}(x/y)$ for all $(x,y)\in \mathbb{R}_+^2$. These computations prove the first inequality in (A) as well as the second equilibrium condition in (C).

To prove the opposite inequality in (A) and the first equilibrium condition in (C) we obtain -- using again that the admissible stopping policy $\tau^\ast=\inf\{t\geq 0: Z_t\not\in (z_{1c}^\ast,z_{2c}^\ast)\}\in \mathcal{T}$ is $\mathbb{Q}^{\theta^{c^\ast}}$-a.s. finite and $\Pi_c(Z_{\tau^\ast})\geq \left(\Pi_{c^\ast}(z_{1c}^\ast)\wedge\Pi_{c^\ast}(z_{2c}^\ast)\right)$ on the set $\tau^{\ast}<\infty$ as well as Lemma \ref{supermartingale} --
\begin{align*}
V_\kappa(x,y) &\geq \inf_{\mathbb{Q}^{\theta}\in \mathcal{P}^\kappa}\mathbb{E}_{\mathbf{x}}^{{\mathbb{Q}^{\theta}}}\left[e^{-r\tau^{\ast}}F(X_{\tau^{\ast}},Y_{\tau^{\ast}})
\mathbbm{1}_{\{\tau^{\ast}<\infty\}}\right]\\
&=\inf_{\mathbb{Q}^{\theta}\in \mathcal{P}^\kappa}\mathbb{E}_{\mathbf{x}}^{{\mathbb{Q}^{\theta}}}\left[e^{-r\tau^{\ast}}X_{\tau^{\ast}}U_{c^\ast}(Z_{\tau^{\ast}})
\Pi_{c^\ast}(Z_{\tau^{\ast}})\mathbbm{1}_{\{\tau^{\ast}<\infty\}}\right]\\
&\geq \left(\Pi_{c^\ast}(z_{1c}^\ast)\wedge\Pi_{c^\ast}(z_{2c}^\ast)\right)\inf_{\mathbb{Q}^{\theta}\in \mathcal{P}^\kappa}\mathbb{E}_{\mathbf{x}}^{{\mathbb{Q}^{\theta}}}\left[e^{-r\tau^{\ast}}X_{\tau^{\ast}}
U_{c^\ast}(Z_{\tau^{\ast}})\mathbbm{1}_{\{\tau^{\ast}<\infty\}}\right]\\
&= \left(\Pi_{c^\ast}(z_{1c}^\ast)\wedge\Pi_{c^\ast}(z_{2c}^\ast)\right)\mathbb{E}_{\mathbf{x}}^{{\mathbb{Q}^{\theta^{c^\ast}}}}
\left[e^{-r\tau^{\ast}}X_{\tau^{\ast}}U_{c^\ast}(Z_{\tau^{\ast}})\mathbbm{1}_{\{\tau^{\ast}<\infty\}}\right]\\
&= \left(\Pi_{c^\ast}(z_{1c}^\ast)\wedge\Pi_{c^\ast}(z_{2c}^\ast)\right)xU_{c^\ast}(z)
\end{align*}
for all $(x,y)\in \{(x,y)\in \mathbb{R}_+^2:z_{1c}^\ast x < y < z_{2c}^\ast x\}$, proving (A). Now, (A) together with the last inequalities in Lemma \ref{supermartingale}, yields the first part of (C).

Finally, noticing that for all $(x,y)\in \{(x,y)\in \mathbb{R}_+^2:z_{1c}^\ast x < y < z_{2c}^\ast x\}$ we have
$$
V_\kappa(x,y)-F(x,y) = xU_{c^\ast}(z)\left(\Pi_{c^\ast}(z_{ic}^\ast)-\Pi_{c^\ast}(z)\right)>0
$$
showing that $\{(x,y)\in \mathbb{R}_+^2:z_{1c}^\ast x < y < z_{2c}^\ast x\}\subseteq C_\kappa$, viz. (B).
\end{proof}

Theorem \ref{t1} states a set of sufficient conditions under which the optimal stopping policy of the considered stopping problem constitutes a two-boundary
stopping policy. Interestingly, according to Theorem \ref{t1}, the pair $(\mathbb{Q}^{{\bm \theta}^{c^\ast}},\tau^\ast)$ constitutes a Nash equilibrium. Hence,
our results show that the decision making problem can be interpreted as a game. The limiting single boundary cases are now summarized in our next two theorems.

\begin{theorem}[upper-boundary case]\label{t3}
	Assume that $2\mu>2\kappa\sigma-\sigma^2$ and that there exists a point $z^\ast\in \argmax\left\{\Pi_{\infty}(z)\right\}\in(0,\infty)$. Then
	\begin{itemize}
		\item[(A)]
		$V_\kappa(x,y) =
		x\Pi_\infty(z^\ast)Q_\kappa(y/x)$ whenever $y>z^\ast x$.
		\item[(B)] If $\Pi_{\infty}(z^\ast)>\Pi_{\infty}(z)$ for all $z>z^\ast$, then $$\{(x,y)\in \mathbb{R}_+^2: y>z^\ast x\}\subseteq C_\kappa:=\{(x,y)\in \mathbb{R}_+^2:V_\kappa(x,y)>F(x,y)\}.$$
		\item[(C)] Let $\tau^\ast\in \mathcal T$ be such that  $\tau^\ast=\inf\{t\geq 0:Z_t<z^\ast\}$ $\mathbb P_{\bf x}$-a.s. for all initial points $(x,y)$ with $y>z^\ast x$. Then, $(\mathbb{Q}^{\kappa},\tau^\ast)$ is a Nash equilibrium in the sense that for all initial points $(x,y)$ with $y < z^\ast x$ it holds that
		\begin{align*} \mathbb{E}_{\mathbf{x}}^{{\mathbb{Q}^{\theta}}}\left[e^{-r\tau^{\ast}}F(X_{\tau^{\ast}},Y_{\tau^{\ast}})\mathbbm{1}_{\{\tau^{\ast}<\infty\}}\right]&\geq \mathbb{E}_{\mathbf{x}}^{\mathbb{Q}^{\kappa}}\left[e^{-r\tau^{\ast}}F(X_{\tau^{\ast}},Y_{\tau^{\ast}})\mathbbm{1}_{\{\tau^{\ast}<\infty\}}\right]\mbox{ for all }\mathbb{Q}^{\theta}\in \mathcal{P}^\kappa\\
		\mathbb{E}_{\mathbf{x}}^{\mathbb{Q}^{\kappa}}\left[e^{-r\tau}F(X_{\tau},Y_{\tau})\mathbbm{1}_{\{\tau<\infty\}}\right]&\leq \mathbb{E}_{\mathbf{x}}^{\mathbb{Q}^{\kappa}}\left[e^{-r\tau^{\ast}}F(X_{\tau^{\ast}},Y_{\tau^{\ast}})\mathbbm{1}_{\{\tau^{\ast}<\infty\}}\right]\mbox{ for all }\tau\in\mathcal T.
		\end{align*}
	\end{itemize}
\end{theorem}
\begin{proof}
Standard computations show that
\begin{align}\label{scale}
\mathbb{Q}_z^\kappa(\tau^\ast<\tau_b)=\frac{\int_{z}^{b}S_\kappa'(t)dt}{\int_{z^\ast}^{b}S_\kappa'(t)dt}=
\frac{\int_{z}^{b}t^{\frac{2(\mu-\kappa\sigma)}{\sigma^2}}e^{\frac{2}{\sigma^2 t}}dt}{\int_{z^\ast}^{b}t^{\frac{2(\mu-\kappa\sigma)}{\sigma^2}}e^{\frac{2}{\sigma^2 t}}dt},
\end{align}
where $\tau_b=\inf\{t\geq 0: Z_t=b\}$ and $z\in(z^\ast,b)$.
Since $\infty$ is a natural boundary for the ratio process, we know that $\lim_{b\rightarrow\infty}\tau_b=\infty$ almost surely. Letting
$b\rightarrow \infty$ and imposing the condition $2\mu>2\kappa\sigma-\sigma^2$ shows that
$$
\mathbb{Q}_z^\kappa(\tau^\ast<\infty)=\lim_{b\rightarrow\infty}\frac{\int_{z}^{b}S_\kappa'(t)dt}{\int_{a}^{b}S_\kappa'(t)dt}=1
$$
proving the $\mathbb{Q}^{\kappa}$-a.s. finiteness of the hitting time $\tau^\ast$ for all initial points satisfying $z>z^\ast$.
The statement holds by a straightforward modification of the arguments in the proof of Theorem \ref{t1}.
\end{proof}

\begin{theorem}[lower-boundary case]\label{t2}
	Assume that  there exists a point $z^\ast\in \argmax\left\{\Pi_{0}(z)\right\}\in(0,\infty)$. Then
	\begin{itemize}
		\item[(A)]
		$V_\kappa(x,y) =
		x\Pi_0(z^\ast)U_0(y/x)$ whenever $y<z^\ast x$.
		\item[(B)] If $\Pi_{0}(z^\ast)>\Pi_{0}(z)$ for all $z<z^\ast$, then $$\{(x,y)\in \mathbb{R}_+^2: y<z^\ast x\}\subseteq C_\kappa:=\{(x,y)\in \mathbb{R}_+^2:V_\kappa(x,y)>F(x,y)\}.$$
		\item[(C)] Let $\tau^\ast\in \mathcal T$ be such that  $\tau^\ast=\inf\{t\geq 0:Z_t>z^\ast\}$ $\mathbb P_{\bf x}$-a.s. for all initial points $(x,y)$ with $y<z^\ast x$. Then, $(\mathbb{Q}^{\theta^{0}},\tau^\ast)$ is a Nash equilibrium in the sense that for all initial points $(x,y)$ with $y < z^\ast x$ it holds that
		\begin{align*} \mathbb{E}_{\mathbf{x}}^{{\mathbb{Q}^{\theta}}}\left[e^{-r\tau^{\ast}}F(X_{\tau^{\ast}},Y_{\tau^{\ast}})\mathbbm{1}_{\{\tau^{\ast}<\infty\}}\right]&\geq \mathbb{E}_{\mathbf{x}}^{\mathbb{Q}^{\theta^{0}}}\left[e^{-r\tau^{\ast}}F(X_{\tau^{\ast}},Y_{\tau^{\ast}})
\mathbbm{1}_{\{\tau^{\ast}<\infty\}}\right]\mbox{ for all }\mathbb{Q}^{\theta}\in \mathcal{P}^\kappa\\
		\mathbb{E}_{\mathbf{x}}^{\mathbb{Q}^{\theta^{0}}}\left[e^{-r\tau}F(X_{\tau},Y_{\tau})\mathbbm{1}_{\{\tau<\infty\}}\right]&\leq \mathbb{E}_{\mathbf{x}}^{\mathbb{Q}^{\theta^{0}}}\left[e^{-r\tau^{\ast}}F(X_{\tau^{\ast}},Y_{\tau^{\ast}})\mathbbm{1}_{\{\tau^{\ast}<\infty\}}\right]
\mbox{ for all }\tau\in\mathcal T.
		\end{align*}
	\end{itemize}
\end{theorem}
\begin{proof}
Making use of the fact that the lower boundary $0$ is entrance and, thus, unattainable for the ratio process $Z$ and the
representation of the hitting time probabilities as in \eqref{scale} shows that $\tau^\ast < \infty$ $\mathbb{Q}^{\theta^{0}}$-a.s. for all initial points $(x,y)$ with $y<z^\ast x$.
The rest follows again from the arguments of Theorem \ref{t1}.
\end{proof}

\section{Explicit Illustration}

\subsection{Integral Option}

We first consider the integral option case $F(x,y)=y = x (y/x)$ considered in \cite{KrMo1994}. It is clear that in this case the exercise payoff grows at most at a unit linear rate and, hence, we can focus on studying the ratio
$$
\Pi_0(z)=\frac{z}{P_\kappa(z)}.
$$
Invoking the finding of part (B) of Lemma \ref{l1} now proves the following.
\begin{theorem}\label{KrMoOpt}
The value of the optimal stopping strategy reads as
\begin{align}
V_\kappa(x,y) = x P_\kappa(y/x)\sup_{u\geq y/x}\left\{\frac{u}{P_\kappa(u)}\right\} =
\begin{cases}
y,&y\geq \bar{z}_\kappa x,\\
x P_\kappa(y/x)\left\{\frac{\bar{z}_\kappa}{P_\kappa(\bar{z}_\kappa)}\right\}, & y < \bar{z}_\kappa x,
\end{cases}
\end{align}
where $\bar{z}_\kappa>1/r$ is the unique root of the first order optimality condition $P_{\kappa}(z)-zP_{\kappa}'(z)=0$. Moreover, the worst case measure is generated by the optimal density generator
$\theta_t^\ast=\kappa\sgn(Z_t-\bar{z}_\kappa)$ for all $t\in \mathbb{R}_+$.
\end{theorem}
\begin{proof}
The alleged value follows directly from part (B) of Lemma \ref{l1} and Theorem \ref{t2}. The optimality of the proposed density generator follows from Lemma \ref{supermartingale}.
\end{proof}
\begin{rmk}
It is worth pointing out that the optimal density generator switches precisely at the optimal exercise boundary from one extreme to another. Thus,
if $y<\bar{z}_\kappa x$, then $\theta_t^\ast=\kappa$ for all $t<\inf\{t\geq 0: Z_t=\bar{z}_\kappa\}$. It is interesting to note that the worst case drift is negative for $Z_t>\bar{z}_\kappa$. It can, however, been shown that the same value comes out when considering $\theta^*_t=\kappa$ for all $t\in \mathbb R_+$, as expected.
\end{rmk}

\begin{figure}[!ht]
\begin{center}
\includegraphics[width=0.6\textwidth]{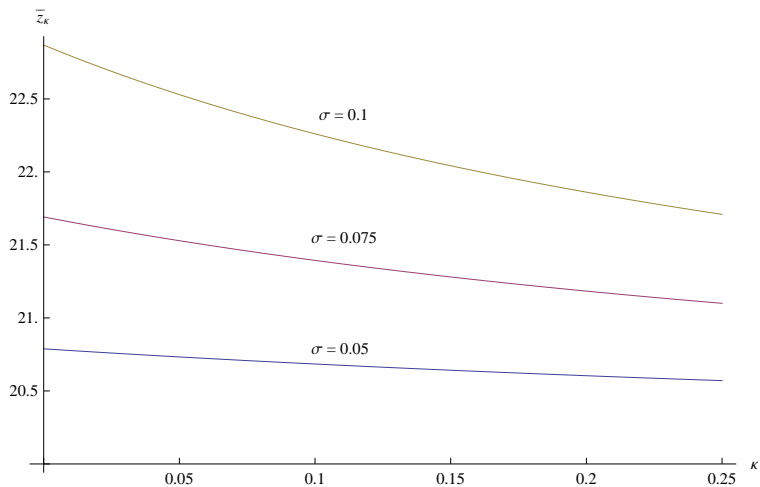}
\caption{\small The optimal exercise boundary $\bar{z}_\kappa$}\label{KrMo}
\end{center}
\end{figure}

The optimal exercise boundary $\bar{z}_\kappa$ is illustrated for three different volatilities $\sigma=5\%,7.5\%,10\%$ in Figure \ref{KrMo} under the assumptions that $\mu = 0.02$ and $r = 0.05$.
As is clear from Figure \ref{KrMo} the optimal boundary is decreasing as a function of the degree of ambiguity illustrating once again the accelerating effect of increased ambiguity on optimal timing. Interestingly, for the chosen parametrization the boundary is an increasing function of volatility. Consequently, the present numerical example supports the view that increased measurable uncertainty (i.e. volatility) decelerates optimal timing by increasing the value of waiting while increased model uncertainty (i.e. ambiguity) has an opposite impact.

\subsection{Exchange Option}

In order to show the usefulness of the developed approach and the sensitivity of the worst case measure with respect to the explicit parametrization of the exercise payoff, consider now the exchange option setting $F(x,y)=(y-K x)^+ = x (y/x-K)^+$, where $K>0$ is a known constant (note that letting $K\downarrow 0$ results in the integral option case treated in the previous subsection).
It is again clear that in this case the exercise payoff $g(z):=(z-K)^+$ grows at most at a unit linear rate. However, since $g'(z)z-g(z)=K>0$ on $(K,\infty)$, we notice that we have to focus on studying the ratio
$$
\Pi_0(z)=\frac{(z-K)^+}{U_0(z)}
$$
also on the set $(\bar{z}_\kappa,\infty)$.
Invoking again the findings of part (B) of Lemma \ref{l1} now results into the following.
\begin{theorem}\label{KrMoOptEx}
The value of the optimal stopping strategy reads as
\begin{align}\label{exchangeval}
V_\kappa(x,y) = x U_0(y/x)\sup_{v\geq y/x}\left\{\frac{(v-K)^+}{U_0(v)}\right\} =
\begin{cases}
y-Kx,&y\geq z^{\ast}_{\kappa} x,\\
x U_0(y/x)\left\{\frac{z^{\ast}_{\kappa}-K}{U_0(z^{\ast}_{\kappa})}\right\}, & y < z^{\ast}_{\kappa} x,
\end{cases}
\end{align}
where $z^{\ast}_{\kappa}>\max(\bar{z}_\kappa,K)$ is the unique root of the first order optimality condition
\begin{align}\label{exchangeopt}
\frac{U_0'(z^{\ast}_{\kappa})K}{S_{-\kappa}'(z^{\ast}_{\kappa})}+
\int_{\bar{z}_\kappa}^{z^{\ast}_{\kappa}}U_0(t)(1+(r-\mu-\kappa\sigma)K-rt)m_{-\kappa}'(t)dt=0.
\end{align}
Moreover, the worst case measure is generated by the optimal density generator
$\theta_t^\ast=\kappa\sgn(Z_t-\bar{z}_\kappa)$ for all $t\in \mathbb{R}_+$.
\end{theorem}
\begin{proof}
Since $z/P_\kappa(z)$ is increasing on $(0,\bar{z}_\kappa)$ we notice that $(z-K)/P_\kappa(z)$ is increasing on $(0,\bar{z}_\kappa)$ as well.
Modifying now slightly the proof of part (B) of Lemma \ref{l1} yields that on $(\bar{z}_\kappa,\infty)$ we have
$$
\frac{U_0(z)-(z-K)U_0'(z)}{S_{-\kappa}'(z)}=\frac{U_0'(z^{\ast}_{\kappa})K}{S_{-\kappa}'(z^{\ast}_{\kappa})}+
\int_{\bar{z}_\kappa}^{z}U_0(t)(1+(r-\mu-\kappa\sigma)K-rt)m_{-\kappa}'(t)dt
$$
implying again that equation $U_0(z)-(z-K)U_0'(z)=0$ has a unique root $z_\kappa^\ast\in(\max(\bar{z}_\kappa,K),\infty)$ so that
$z_\kappa^\ast$ constitutes the global maximum point of $(z-K)/U_0(z)$. The alleged representation \ref{exchangeval} then follows from
Theorem \ref{t2}. The proposed optimal density generator then follows from Lemma \ref{supermartingale}.
\end{proof}

Theorem \ref{KrMoOptEx} demonstrates how sensitive the value and the associated worst case measure are with respect to the parametric specification of the payoff even in the linear case. In the extreme case where $K=0$ the optimal density generator switches only at the optimal exercise. However, as soon as the multiplier satisfies the inequality $K>0$ the density generator switches prior exercise provided that the initial state $z$ is below $\bar{z}_\kappa$.

\begin{figure}[!ht]
\begin{center}
\includegraphics[width=0.6\textwidth]{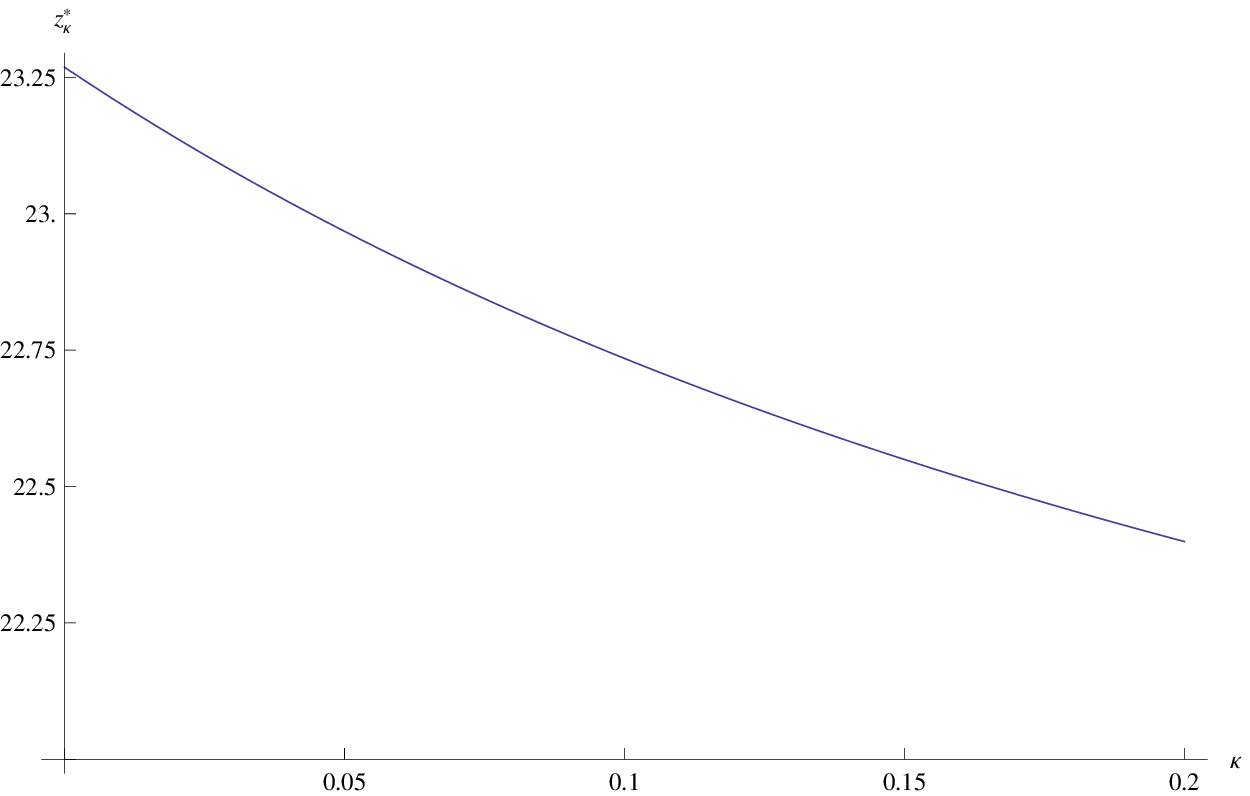}
\caption{\small The optimal exercise boundary $z^\ast_\kappa$}\label{KrMoExch}
\end{center}
\end{figure}

The optimal exercise boundary $z^\ast_\kappa$ is illustrated in Figure \ref{KrMoExch} under the assumptions that $\mu = 0.02,\sigma=0.1,K=0.5,$ and $r = 0.05$.
In line with the findings in our previous section, Figure \ref{KrMoExch} again indicates that increased ambiguity accelerate timing by decreasing the optimal stopping boundary. Interestingly, for the chosen parametrization the boundary is an increasing function of volatility. Consequently, the present numerical example supports the view that increased measurable uncertainty (i.e. volatility) decelerates optimal timing by increasing the value of waiting while increased model uncertainty (i.e. ambiguity) has an opposite impact.

\subsection{Floor Option}

In order to illustrate our general results we focus now on an explicitly parameterized floor option example $F(x,y)=\max(x,y)=x\max(1,y/x)$.
Given the boundary behavior of the underlying at the lower boundary $0$, the limit $\lim_{z\rightarrow 0+}P_{\kappa}(z)=1$ and the fact that the exercise payoff grows at most unit linear rate, we first consider the ratio
$$
R(z)=\frac{z}{U_0(z)}.
$$
As proved in part (B) of Lemma \ref{l1}, there is a unique maximizer $\bar{z}_\kappa=\argmax\{z/P_\kappa(z)\}>1/r$ satisfying the ordinary first order condition
$\bar{z}_\kappa P_\kappa'(\bar{z}_\kappa)=P_\kappa(\bar{z}_\kappa)$. If $\bar{z}_\kappa\geq P_\kappa(\bar{z}_\kappa)$, then the monotonically increasing and continuously differentiable function
$$
\tilde{V}_\kappa(x,y)=\begin{cases}
y,&y\geq \bar{z}_\kappa x,\\
xR(\bar{z}_\kappa)P_\kappa(y/x),&y<\bar{z}_\kappa x
\end{cases}
$$
dominates the exercise payoff $\max(x,y)$ for all $(x,y)\in\mathbb{R}_+^2$. Moreover, $\tilde{V}_\kappa(x,y)=V_\kappa(x,y)$ follows from Theorem \ref{t2}.

If, however, $\bar{z}_\kappa < P_\kappa(\bar{z}_\kappa)$, then the problem becomes a two boundary problem. Utilizing the monotonicity properties of the function $U_c(z)$ shows that $1/U_c(z)$ attains its maximum at $c$ implying that the lower boundary coincides with the reference point $c$. On the other hand, since
$$
\frac{d}{dz}\frac{z}{U_c(z)}=\frac{U_c(z)-U_c'(z)z}{U_c^2(z)}\gtreqqless 0,\quad z\lesseqqgtr \hat{z}_c,
$$
we notice that the ratio $z/U_c(z)$ is maximized at the threshold $\hat{z}_c>c$ satisfying the condition \eqref{switch}. Since
$$
\frac{1}{U_{c^\ast}(z_1^\ast)}=\frac{z_2^\ast}{U_{c^\ast}(z_2^\ast)},
$$
$z_1^\ast = c^\ast$, and $z_2^\ast=\hat{z}_{c^\ast}$, we notice that
the critical reference point $c^\ast$ satisfies the equation
$$
\frac{P_{\kappa}'(c^\ast)}{S_\kappa'(c^\ast)}Q_\kappa(\hat{z}_{c^\ast})-\frac{Q_{\kappa}'(c^\ast)}{S_\kappa'(c^\ast)}P_\kappa(\hat{z}_{c^\ast})=B_{\kappa}\hat{z}_{c^\ast}.
$$
\begin{rmk}
It is worth noticing that the considered floor option problem could alternatively be obtained by considering directly the boundary value problem
\begin{align*}
&\frac{1}{2}\sigma^2z^2h''(z) + (1-\mu z+\kappa\sigma z)h'(z)-(r-\mu+\kappa\sigma)h(z)=0,\\
& h(z_1^\ast)=1, h'(z_1^\ast)=0,\\
& h(z_2^\ast)=z_2^\ast, h'(z_2^\ast)=1.
\end{align*}
\end{rmk}

\begin{figure}[!ht]
\begin{center}
\includegraphics[scale=0.75]{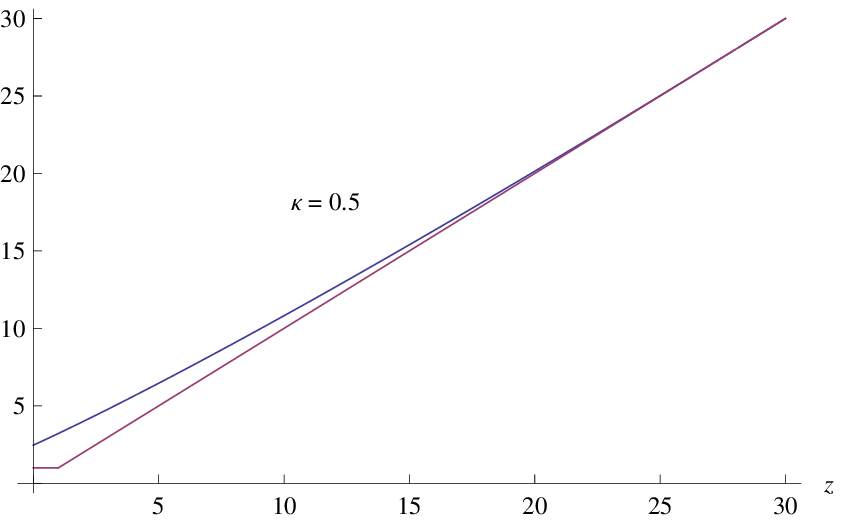}
\includegraphics[scale=0.75]{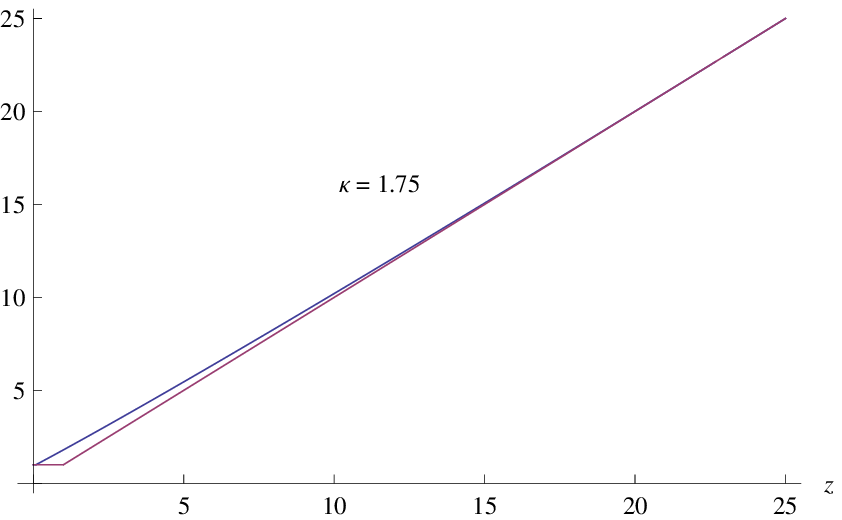}
\caption{\small The value of the optimal stopping strategy}\label{Floorfig}
\end{center}
\end{figure}

We illustrate our results numerically as functions of the degree of ambiguity $\kappa$ by assuming that $\mu = 0, r = 0.05$, and $\sigma = 0.5$.
In this case, the critical degree of ambiguity $\kappa$ at which the problem becomes a two-boundary problem is $\hat{\kappa}\approx1.59795$. The value functions for $\kappa=0.5,1.75$ are illustrated in Figure \ref{Floorfig} ($\bar{z}_{0.5}\approx 27.9912$, and $(z_1^\ast,z_2^\ast)\approx(0.0854, 22.6858)$ for $\kappa=1.75$).
As is clear from Figure \ref{Floorfig}, in order to end up with a two boundary setting the parameter values have to be unrealistically large and, hence, from a practical point of view the single boundary setting is more prevalent.

\section{Concluding Comments}
We analyzed the optimal timing policy of an ambiguity averse decision maker facing Knightian uncertainty in a two-dimensional setting based on an ordinary geometric Brownian motion and its integral process. By focusing on measurable and positively homogeneous exercise payoffs, we delineated circumstances under which the considered stopping problem as well as the worst case measure can be solved explicitly and the resulting solution constitutes a Nash equilibrium. There are naturally several directions towards which our analysis could be extended. To some extent the most natural direction would be to consider more general payoff structures than just positively homogeneous ones. Unfortunately, such extension results easily in situations where dimensionality reduction no longer applies and solving the considered stopping problem requires the analysis of a very challenging second order partial differential equation. A second natural direction would be to consider mean reverting diffusions and investigating whether our main conclusions would remain unchanged at least qualitatively as long as the exercise payoff remains positively homogeneous. A third direction would be to increase the dimensionality of the factor dynamics driving the uncertainty in the considered class of models in the spirit of \cite{AlCh2019}. Such an extension would probably result in the addition of multiple switching points for the optimal density generators characterizing the worst case measure. In light of the relative complexity of the considered problem even in a single factor setting, such an extension would most likely require the analysis of a problem where the family of excessive mappings needed for the characterization of the optimal policy is not parameterized by a single reference point but by several endogenously determined reference points. Since all these quantities have to be solved simultaneously from a complex set of equations, we leave that problem for the future.

\bibliographystyle{apalike}
\bibliography{Knight}

\end{document}